# Upgrading the Detection of Electrocatalyst Degradation During the Oxygen Evolution Reaction

*Marcel Risch*

*Nachwuchsgruppe Gestaltung des Sauerstoffentwicklungsmechanismus*

*Helmholtz-Zentrum Berlin für Materialien und Energie GmbH*

*Hahn-Meitner-Platz 1, 14019 Berlin*

Corresponding author: Dr. Marcel Risch (marcel.risch@helmholtz-berlin.de)

**Abstract**

Electrocatalysts for the oxygen evolution reaction (OER) are an important component for the transition from fossil to sustainable energy. Commercialization of cost-effective earth-abundant electrocatalysts is in large parts hindered by their degradation. In this short review, I identify common processes leading to a decrease in electrocatalyst activity, followed by an introduction of staple methods to determine degradation electrochemically and by additional physical characterization, which has the potential to remove ambiguities of purely electrochemical studies. I conclude by a summary of the key challenges for an accurate determination of degradation processes and highlight interesting directions to advance the understanding of degradation processes on electrocatalysts.

**Introduction**

Electrocatalysts for the oxygen evolution reaction (OER) are an important component for the transition from fossil to sustainable energy as they provide the protons for the electrocatalytic production of $H_2$ and other valuable products.[1] Degradation processes are a key aspect that hinders the commercialization of OER electrocatalysts, particularly when made from earth-abundant elements. The mechanisms of electrocatalyst degradation have been reviewed before in detail for atomistic processes on electrocatalyst materials [2–4] and membrane electrode assemblies (MEA)[3,4] used in devices.

In this short review, I comment on the detection of electrocatalyst degradation during the OER. Firstly, I identify common processes leading to a decrease in electrocatalyst activity on various length scales, where I point out that only a subset constitutes electrocatalyst degradation. The misassignment is in part due to the widespread determination of degradation solely by electrochemical methods, which I argue is insufficient. This is followed by a brief discussion of suitable complementary methods to corroborate and identify the degradation process(es). I conclude by a summary of the key challenges for an accurate determination of degradation processes and highlight interesting directions to advance the understanding of degradation processes on electrocatalysts.

**What processes degrade the apparent electrocatalytic activity?**

Several processes may lead to an actual or apparent degradation of the electrocatalyst (Fig. 1). Thus, it is important to identify the origins of electrocatalyst degradation to device appropriate measures for quantification of the loss or in the best-case mitigation thereof. Spöri et al.[4] categorized catalyst stability in material stability and operational stability. I take a related but different view to categorize possible processes as readily reversible processes on the liquid-solid interface (bubbles, adsorption in Fig. 1a) or increasingly irreversible changes of the electrocatalyst materials and its support (Fig. 1b-d). I consider only the latter electrocatalyst degradation whereas the former could be called operational degradation as it is usually not detectable after electrocatalysis.

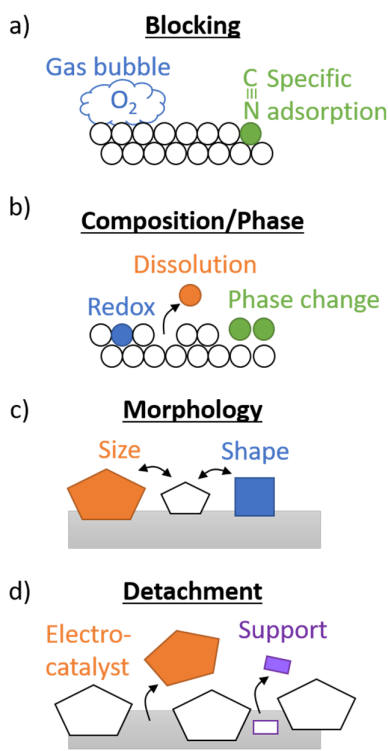

**Figure 1.** Processes leading to an apparent degradation of electrocatalyst activity and/or selectivity: (a) blocking of active sites, (b) changes of composition and/or phase, (c) changes of morphology, and (d) detachment of the electrocatalyst and support.

Possible active sites on the surface of the electrocatalyst may be blocked over time without degrading the electrocatalyst material. Typical examples are specifically adsorbed anions from the electrolyte such as phosphate, which reduce the number of possible active sites.[5,6] The electrolyte cations also affect the measured activities. The cations may interact with reaction intermediates[7] or alter the properties of the double layer such that the OER decreases in the presence of large cations.[8] Cations such as $Na^+$ are released from typical laboratory glass,[9] which may create a reduction of activity over time. In general, the effect of the electrolyte of intentionally or unintentionally electrolyte ions on the OER is not well studied and these effects may be frequently attributed to a materials change.

Gas bubbles on the surface may also prevent the reactants of the OER (hydroxide or water[10]) from reaching active sites. They can be introduced by gas purging, immersing electrodes or be produced in situ by the OER where the bubble release depends, e.g., on the electrocatalyst morphology.[11] El-Sayed et al.[12] pointed out that microbubbles result in reduced currents over time, which may be mistaken for electrocatalyst degradation. Rotation and bath sonication was insufficient to prevent the apparent degradation.[13] In summary, blockage of active sites is expected to occur frequently due to non-optimal experimental designs. Improved experiments can mitigate or even avoid the reduction of the measured currents with time for blocking processes.

Electrocatalysts frequently undergo changes of their composition and structure on the atomic scale during the OER (Fig. 1b). The loss of the active site to the electrolyte is an obvious example leading to electrocatalyst degradation by dissolution. Yet, neighboring atoms or redox-inert cations may also dissolve and thereby degrade the properties of the active sites remaining in the solid phase. Note that dissolution of catalytically inert elements may have no effect until their number is sufficiently low to trigger other materials changes such as a phase change (see below). Furthermore, the active sites of the electrocatalyst can undergo redox reactions that render them inactive, e.g., the oxidation of $Mn^{3+}$ to $Mn^{4+}$ at potentials lower than the onset of the OER.[14,15] Other redox changes can drastically reduce the conductivity of the catalyst material.[16] Dissolution and/or metal redox reactions may also lead to phase changes of the electrocatalyst material.[3] This is perhaps best studied for NiOOH where the formation of the so-called overcharged γ-NiOOH phase constitutes catalyst degradation.[17,18]

Dissolution, redox chemistry and phase changes are commonly summarized as corrosion, which the IUPAC defines as "an irreversible interfacial reaction of a material (metal, ceramic, polymer) with its environment which results in consumption of the material or in dissolution into the material of a component of the environment."[19] Electrocatalyst degradation is most frequently attributed to corrosion, which can be understood and modeled atomistically.[2] Note that the atoms of an electrocatalyst material are not immobile under typical electrochemical conditions and their location may oscillate over short time scales, while these changes average out on longer time scales. We are just uncovering these effects with high relevance for degradation though new methods such as environmental transmission electron microscopy (ETEM) studies.[20]

Electrocatalyst degradation may also occur on larger scales, namely when the morphology changes (Fig. 1c) or when macroscopic parts of the electrode detach (Fig. 1d).[4] Morphology changes include the size and shape of the electrocatalyst particles. Both is expected to alter the distribution of surface facets, where less degradation of the (100) surface of Co spinels as compared to the (111) surface has been attributed to the lower surface energy of the former.[21] It may be possible to normalize for the change in particle size, which increases the surface area for unchanged mass loading (i.e., negligible dissolution and detachment) but the needed measurements for surface normalization are rarely performed in situ with sufficient frequency to correct for the effect.

The effect of detachment (or erosion) is similar to dissolution in the sense that active sites are lost (Fig. 1d). The electrocatalyst particles could detach as one piece or parts may detach. Another possibility is the detachment of the support material which may impact the function of the support such as electric connection of the electrocatalyst particles. (Note that the function of the support may also be affected by corrosion[22]). Additionally, the detachment of the support is coupled to loss of electrocatalyst material that may be dispersed on the support or mechanically fixed to it. It is well known that carbon supports corrode at potentials lower than the onset of the OER.[23,24] Non-optimized electrocatalyst electrodes may also detach as a single film from their support, i.e. delaminate.[4] In summary, degradation of the composite electrode or MEA, i.e. electrocatalyst, support and potential additives, may happen frequently. Yet, degradation on the relevant scale is understudied.

In conclusion, multiple processes are relevant to fully understand the observed reduction in currents over time and to associate an appropriate electrocatalyst degradation process. Often, multiple processes occur simultaneously and/or sequentially, which imposes high demands on experimental design to isolate the effects of a single processes. I have also highlighted that a comprehensive picture of electrocatalyst degradation requires investigations on a wide range of lengths scales from the atomic to macroscopic (electrode areas 0.1 - 1 $cm^2$ in academic research and up to 100 $cm^2$ for MEA).

**How is electrocatalyst degradation detected?**

In most studies, electrocatalyst degradation is detected as the by-product of another electrochemical experiment such as a protocol to determine activity or repeated applications of the activity protocol rather than an experiment designed to decouple the processes illustrated in Fig. 1. Examples for a degradation-specific protocol are the accelerated degradation testing (AST) protocols proposed by Spöri et al.[4,25] Popular methods for degradation studies are cyclic voltammetry (CV), chronoamperometry (CA) or chronopotentiometry (CP).

In CV (Fig. 2a), the potential is cycled between appropriate limits where the upper limit needs to include the current rise due to the OER. Often the cycling is performed faster as compared to activity studies (e.g., 100 mV/s vs. 10 mV/s) and up to several 10,000's cycles are recorded.[26] Often, new processes become apparent not only during the first 100 cycles[13,27] but also during much later cycles.[28] The current density at one or more reference potentials is then plotted against cycling, ideally after correction for capacitive and ohmic currents.[29] Possible degradation is seen as a reduction of the currents with the cycle number, either in absolute values or relative to the initial (or another relevant) current density.

In CA (Fig. 2b), the electrode is held at a selected reference potential, e.g., 0.27 V overpotential,[4] and the current density is recorded. Possible degradation is seen as a reduction of the current with time. Usually, durations of a few hours are reported (e.g. ref. [30]), which is short for a degradation test (see next paragraph). Absolute or relative changes of the current can be calculated from the data. Often, bar charts are often found to compare electrocatalysts.

In CP (Fig. 2c), the electrode is held at a selected reference current density such as 10 mA/cm$^2$ [31] and the measured potential is recorded. CP is used more frequently than CA in electrochemical degradation studies. Possible degradation is seen as an increase in (over)potential with time. Two hours are seen frequently as used in the important benchmarking work of McCrory et al.[31] However, two hours are only a first step for screening and should be complemented by longer experiments, where tests for longer than a week (e.g. 1000 h ~ 6 weeks [32]) are unfortunately still rare. For this data, bar graphs are common and also a plot of the overpotential at the start of the experiments vs. the end is often found. In the latter plot, degradation is indicated by data above a line with unit slope.

The degree of current reduction or (over)potential increase can be quantified in all of these experiments, which is unfortunately rarely done. I need to point out that the duration of the experiment is only a measure of the electrocatalyst degradation when the experiment is stopped by complete deactivation of the electrocatalyst. In all other cases, absolute or relative changes of the current density or (over)potential should be used to compare degradation of electrocatalysts in a specified time interval.

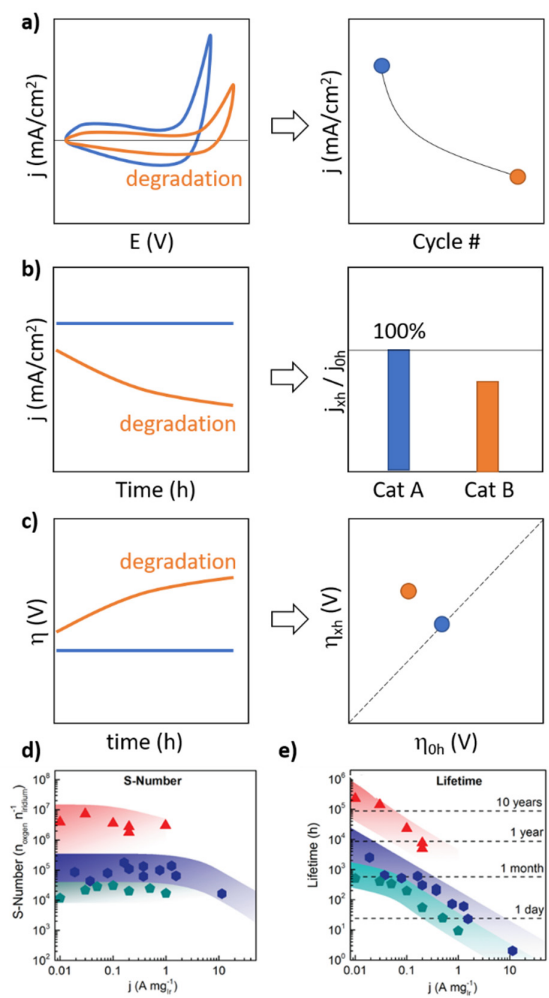

**Figure 2.** Possible ways to pre-screen for electrocatalyst degradation by electrochemical methods and exemplary quantitative representation: (a) Cyclic voltammetry and trends of current at a reference potential with cycling; (b) Chronoamperometry at a selected overpotential and bar plot of relative changes in current density during the measurement; (c) Chronopotentiometry at a selected current density and plot of the overpotential at the end of the measurement ($\eta_{xh}$) relative to that at the beginning ($\eta_{0h}$). (d) The S-number[33] for various Ir oxides and (e) estimated catalyst lifetimes. Reprinted by permission from Springer Nature, Nature Catalysis, ref. [33], Copyright 2018.

The activity-stability factor (ASF)[34] and stability number (S-number)[33] go one step further in the quantification of degradation. The latter S-number also allows estimating the lifetime of the electrocatalyst (Fig. 2d-e). It is similar to the turn-over number (TON) in homogenous (electro)catalysis and equals the ratio of the number of oxygen molecules that can be evolved per estimated active site. A quantification of the dissolved ion(s) is required for the analysis, which is not readily available so that it is less commonly reported as compared to the purely electrochemical quantifications. The discussion of the S-number supports the need of complementing electrochemical data with further investigations to unambiguously identify electrocatalyst degradation and then quantify it.

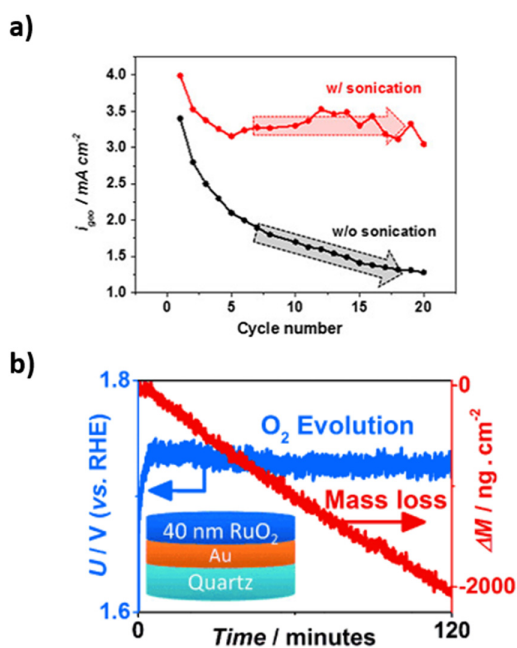

**Figure 3.** Two important insights for understanding the degradation of electrocatalysts for the oxygen evolution reaction: (a) a reduction in current density may not be due to irreversible materials changes but rather due to bubbles blocking the electrocatalyst surface. Reproduced with permission from ref. [13], Copyright 2020 American Chemical Society. (b) the absence of changes in electrocatalytic studies does not always indicate the absence of electrocatalyst degradation. Reproduced from ref. [35], Copyright 2014, Wiley and Sons.

To further illustrate the ambiguity of electrochemical data, I highlight the results of two key studies in Fig. 3. The aforementioned studies of El-Sayed et al.[12,13] show an apparent decrease in current density in a CV experiment using a rotating disk electrode (RDE). Sonicating the electrolyte to remove bubbles (or waiting at open-circuit; not shown) partially mitigates the current loss (Fig. 3a). On the other hand, Frydendal et al.[35] studied $RuO_2$ and $MnO_2$ using an electrochemical quartz microbalance. While there was no change in (over)potential that would indicate degradation, the mass loss clearly indicated loss of the electrocatalyst material. Therefore, a change in (over)potential or current density does not necessarily indicate electrocatalyst degradation and electrocatalyst degradation is possible in the absence of detectable changes in current density or (over)potential. Thus, the two selected studies substantiate the need to complement electrochemical studies of degradation with gravimetric, microscopic, spectroscopic or other suitable methods.

Examples of direct and indirect methods to investigate the processes in Fig. 1 are summarized in Table 1. Direct methods measure the existence of some species involved in the process. Examples are the detection of adsorbates by their fluorescence or atomic force microscopy (AFM) and the determination of a change in the redox state by X-ray photoelectron spectroscopy (XPS) or X-ray absorption spectroscopy (XAS). Indirect methods vary a key parameter affecting the process such as the concentration of a species that could specifically adsorb or replacing a redox-active ion with an inactive one. Table 1 is by no means comprehensive but should serve as a starting point to identify suitable methods to identify and/or study the processes responsible for changes in current density or (over)potential to understand degradation of the sample of interest. Further discussion of detection methods and characterization examples may also be found in refs.[2–4]

**Table 1.** Processes leading to a change in current density and (over)potential and suggested exemplary methods to directly or indirectly study the process.

| Process | Selected methods and approaches |
|---|---|
| Gas bubble | Direct: spatially resolved $O_2$ fluorescence [36] |
| | Indirect: Electrochemical protocol with resting times and sonication [13] |
| Specific adsorption | Direct: in situ AFM[a] [37] |
| | Indirect: concentration variation of expected adsorbing ion [38] |
| Dissolution | Direct: ICP-MS[b] on electrolyte;[33] EDS/EDX[c] on degraded material [39] |
| Redox | Direct: XAS[d], [15] XPS[e],[40] titration [41] |
| | Indirect: Replacing a redox-active ion by a redox-inactive ion during synthesis [42] |
| Phase Change | Direct: XRD[f],[28] analysis of the EXAFS[g],[28] |
| Morphology | Direct: AFM[a],[43], TEM[h],[44] SEM[i] [28] |
| Detachment | Direct: IL-TEM[j] [40] |

[a] atomic force microscopy; [b] inductively coupled plasma mass spectrometry; [c] energy-dispersive X-ray spectroscopy; [d] X-ray absorption spectroscopy; [e] X-ray photoelectron spectroscopy; [f] X-ray diffraction; [g] extended X-ray absorption fine structure; [i] transmission electron microscopy; [j] scanning electron microscopy; [k] identical location TEM.

**Conclusion and Outlook**

In the last decade, more weight has been put on aspects of electrocatalyst degradation in studies of the OER; virtually all studies with a focus on electrocatalyst materials published nowadays contain data to discuss electrocatalyst degradation. Including these discussions is an auspicious direction for the field to identify promising electrocatalyst materials for device tests. The sorest spots hindering a better understanding degradation processes are:

- Degradation may happen on very different length scales. While there is much attention on atomistic processes, which are certainly very important, the processes on larger length scales such as blocking by bubble formation and particle/support detachment are clearly understudied relative to their importance for use in devices.
- Changes of the electrocatalyst material need to be clearly distinguished from those due to blockage of surface sites to device appropriate strategies for remediation of the degradation process. Many electrocatalysts were likely discarded prematurely due to inappropriate electrolyte composition or experimental apparatuses.
- The identification of the suspected process of degradation can likely be achieved in a well-designed post-mortem investigation with the caveat that the degradation of the material may occur before electrocatalysis,[45] e.g., a reaction with an additive during sample preparation.[46] Understanding the mechanism of degradation needs an operando or in situ experiment, particularly those with long duration.[45] Mechanistic insight is a key prerequisite for the knowledge-guided mitigation of electrocatalyst degradation, e.g. by optimizing the composition.[47]
- The elucidation of degradation has higher demands on experimental design as compared to activity (benchmarking) studies simply due to the needed long duration, during which important parameters, such as temperature and electrolyte concentration, need to be held constant. Optimizing the experimental design is particularly important for another positive recent direction of the field, namely the

- investigations of "industry-relevant" conditions of high currents, high electrolyte concentration and temperature above room temperature.
- There is no common protocol (or industry standard) used to quantify the degree of electrocatalyst degradation, which is a shortcoming shared with activity benchmarking studies. This reduces the comparability among studies. I advocate using a specific protocol for degradation studies such as the one in ref. [4] as degradation studies have specific experimental demands differing from activity benchmarking studies (see above).
- There are also no minimum requirements for reporting electrochemical degradation data. I plead to thrive for a quantitative report that can be compared to other works. For this, one needs to report both absolute current density and (over)potential differences as well as relative changes in current density for clearly specified conditions during multiple defined durations, e.g., 2 h (reference data in refs. [31,48,49]), 24 h, 1 week, 1 month, etc. Meaningful degradation studies need to be performed at the same current density or (over)potential used for activity studies. For solar fuels devices 10 mA/cm$^2_{geo}$[31] or 400 mV overpotential.[50] For electrolyzers, Spöri et al.[4] proposed 270 mV overpotential and 10 A/g [51] for material testing and 2 A/cm$^2$ for polymer electrolyte membrane (PEM) electrolyzers, while I note that 0.5 A/cm$^2$ are typical for alkaline electrolyzers. The field urgently needs quantitative data (even with expected high scatter due to different protocols) to identify the most promising materials families and/or morphologies to focus on for the next generation of durable electrocatalysts, particularly those made from non-critical and abundant materials.
- I have focused on degradation of structural stability in this review. Yet, there is also recent work on dynamic stability, [52] where the degradation processes are reversed under specific conditions, which leads to self-repair or even self-healing. Dynamic stability likely only affects changes in atomic composition and structure, where it is an attractive alternative to structural stability that would lead to no detectable degradation by the methods described herein.

In this short review, I summarized some existing shortcomings in the determination of electrocatalyst degradation during the OER and commented which aspects should be urgently upgraded. I wish that my thoughts stipulate discussion in the electrocatalyst community how to make electrocatalyst characterization, in particular of degradation, more reproducible and comparable. At the same time, I hope that this review provides clear guidance to newcomers from other fields how to report (ideally the lack of) degradation of new electrode materials.

**Competing interest**

The author does not declare a competing interest

**Acknowledgement**

This project has received funding from the European Research Council (ERC) under the European Union's Horizon 2020 research and innovation programme under grant agreement No. 804092.

**References**

[1] M. Risch, Greater than the sum of its parts, Nat. Energy. 6 (2021) 576–577. https://doi.org/https://doi.org/10.1038/s41560-021-00850-5.

[2] F.-M. Li, L. Huang, S. Zaman, W. Guo, H. Liu, X. Guo, B.Y. Xia, Corrosion Chemistry

of Electrocatalysts, Adv. Mater. (2022) 2200840. https://doi.org/https://doi.org/10.1002/adma.202200840.

[3]   F.-Y. Chen, Z.-Y. Wu, Z. Adler, H. Wang, Stability challenges of electrocatalytic oxygen evolution reaction: From mechanistic understanding to reactor design, Joule. 5 (2021) 1704–1731. https://doi.org/https://doi.org/10.1016/j.joule.2021.05.005.

[4]   C. Spöri, J.T.H. Kwan, A. Bonakdarpour, D.P. Wilkinson, P. Strasser, The Stability Challenges of Oxygen Evolving Catalysts: Towards a Common Fundamental Understanding and Mitigation of Catalyst Degradation, Angew. Chemie Int. Ed. 56 (2017) 5994–6021. https://doi.org/10.1002/anie.201608601.

[5]   K. Iizuka, T. Kumeda, K. Suzuki, H. Tajiri, O. Sakata, N. Hoshi, M. Nakamura, Tailoring the active site for the oxygen evolution reaction on a Pt electrode, Commun. Chem. 5 (2022) 126. https://doi.org/10.1038/s42004-022-00748-7.

[6]   H. Dau, C. Limberg, T. Reier, M. Risch, S. Roggan, P. Strasser, The Mechanism of Water Oxidation: From Electrolysis via Homogeneous to Biological Catalysis, ChemCatChem. 2 (2010) 724–761. https://doi.org/10.1002/cctc.201000126.

[7]   A.C. Garcia, T. Touzalin, C. Nieuwland, N. Perini, M.T.M. Koper, Enhancement of Oxygen Evolution Activity of Nickel Oxyhydroxide by Electrolyte Alkali Cations, Angew. Chemie Int. Ed. 58 (2019) 12999–13003. https://doi.org/https://doi.org/10.1002/anie.201905501.

[8]   J. Huang, M. Li, M.J. Eslamibidgoli, M. Eikerling, A. Groß, Cation Overcrowding Effect on the Oxygen Evolution Reaction, JACS Au. 1 (2021) 1752–1765. https://doi.org/10.1021/jacsau.1c00315.

[9]   D.J. Backhouse, A.J. Fisher, J.J. Neeway, C.L. Corkhill, N.C. Hyatt, R.J. Hand, Corrosion of the International Simple Glass under acidic to hyperalkaline conditions, Npj Mater. Degrad. 2 (2018) 29. https://doi.org/10.1038/s41529-018-0050-5.

[10]   S. Haghighat, J.M. Dawlaty, Continuous Representation of the Proton and Electron Kinetic Parameters in the pH–Potential Space for Water Oxidation on Hematite, J. Phys. Chem. C. 119 (2015) 6619–6625. https://doi.org/10.1021/acs.jpcc.5b00053.

[11]   A.R. Zeradjanin, A.A. Topalov, Q. Van Overmeere, S. Cherevko, X. Chen, E. Ventosa, W. Schuhmann, K.J.J. Mayrhofer, Rational design of the electrode morphology for oxygen evolution-enhancing the performance for catalytic water oxidation, RSC Adv. 4 (2014) 9579–9587. https://doi.org/10.1039/c3ra45998e.

[12]   H.A. El-Sayed, A. Weiß, L.F. Olbrich, G.P. Putro, H.A. Gasteiger, OER Catalyst Stability Investigation Using RDE Technique: A Stability Measure or an Artifact?, J. Electrochem. Soc. 166 (2019) F458–F464. https://doi.org/10.1149/2.0301908jes.

[13]   A. Hartig-Weiss, M.F. Tovini, H.A. Gasteiger, H.A. El-Sayed, OEr catalyst durability tests using the rotating disk electrode technique: The reason why this leads to erroneous conclusions, ACS Appl. Energy Mater. 3 (2020) 10323–10327. https://doi.org/10.1021/ACSAEM.0C01944.

[14]   M. Risch, K.A. Stoerzinger, B. Han, T.Z. Regier, D. Peak, S.Y. Sayed, C. Wei, Z.J. Xu, Y. Shao-Horn, Redox Processes of Manganese Oxide in Catalyzing Oxygen Evolution and Reduction: An In Situ Soft X-ray Absorption Spectroscopy Study, J. Phys. Chem. C. 121 (2017) 17682–17692. https://doi.org/10.1021/acs.jpcc.7b05592.

[15]   M. Baumung, L. Kollenbach, L. Xi, M. Risch, Undesired Bulk Oxidation of

LiMn$_2$O$_4$ Increases Overpotential of Electrocatalytic Water Oxidation in Lithium Hydroxide Electrolytes, ChemPhysChem. 20 (2019). https://doi.org/10.1002/cphc.201900601.

[16] M.B. Stevens, L.J. Enman, A.S. Batchellor, M.R. Cosby, A.E. Vise, C.D.M. Trang, S.W. Boettcher, Measurement Techniques for the Study of Thin Film Heterogeneous Water Oxidation Electrocatalysts, Chem. Mater. 29 (2017) 120–140. https://doi.org/10.1021/acs.chemmater.6b02796.

[17] A. Mavrič, M. Fanetti, Y. Lin, M. Valant, C. Cui, Spectroelectrochemical tracking of nickel hydroxide reveals its irreversible redox states upon operation at high current density, ACS Catal. 10 (2020) 9451–9457. https://doi.org/10.1021/acscatal.0c01813.

[18] F. Dionigi, P. Strasser, NiFe-Based (Oxy)hydroxide Catalysts for Oxygen Evolution Reaction in Non-Acidic Electrolytes, Adv. Energy Mater. 6 (2016) 1600621. https://doi.org/10.1002/aenm.201600621.

[19] IUPAC, Compendium of Chemical Terminology , 2nd ed. (the "Gold Book"), Blackwell Scientific Publications, Oxford, 1997. https://doi.org/10.1351/goldbook.

[20] G. Lole, V. Roddatis, U. Ross, M. Risch, T. Meyer, L. Rump, J. Geppert, G. Wartner, P. Blöchl, C. Jooss, Dynamic observation of manganese adatom mobility at perovskite oxide catalyst interfaces with water, Commun. Mater. 1 (2020) 68. https://doi.org/10.1038/s43246-020-00070-6.

[21] Z. Chen, C. Kronawitter, B. Koel, Facet-dependent activity and stability of Co3O4 nanocrystals towards the oxygen evolution reaction, Phys. Chem. Chem. Phys. (2015). https://doi.org/10.1039/C5CP02876K.

[22] B. Han, M. Risch, S. Belden, S. Lee, D. Bayer, E. Mutoro, Y. Shao-Horn, Screening Oxide Support Materials for OER Catalysts in Acid, J. Electrochem. Soc. (2018). https://doi.org/10.1149/2.0921810jes.

[23] S.G. Ji, H. Kim, W.H. Lee, H.-S. Oh, C.H. Choi, Real-time monitoring of electrochemical carbon corrosion in alkaline media, J. Mater. Chem. A. 9 (2021) 19834–19839. https://doi.org/10.1039/D1TA01748A.

[24] S.J. Ashton, M. Arenz, Comparative DEMS study on the electrochemical oxidation of carbon blacks, J. Power Sources. 217 (2012) 392–399. https://doi.org/10.1016/j.jpowsour.2012.06.015.

[25] C. Spöri, C. Brand, M. Kroschel, P. Strasser, Accelerated Degradation Protocols for Iridium-Based Oxygen Evolving Catalysts in Water Splitting Devices, J. Electrochem. Soc. 168 (2021) 34508. https://doi.org/10.1149/1945-7111/abeb61.

[26] C. Spöri, J.T.H. Kwan, A. Bonakdarpour, D.P. Wilkinson, P. Strasser, The Stability Challenges of Oxygen Evolving Catalysts: Towards a Common Fundamental Understanding and Mitigation of Catalyst Degradation, Angew. Chemie Int. Ed. 56 (2017) 5994–6021. https://doi.org/10.1002/anie.201608601.

[27] J. Villalobos, R. Golnak, L. Xi, G. Schuck, M. Risch, Reversible and irreversible processes during cyclic voltammetry of an electrodeposited manganese oxide as catalyst for the oxygen evolution reaction, J. Phys. Energy. 2 (2020) 034009. https://doi.org/10.1088/2515-7655/ab9fe2.

[28] J. Villalobos, D. González-Flores, R. Urcuyo, M.L. Montero, G. Schuck, P. Beyer, M. Risch, Requirements for Beneficial Electrochemical Restructuring: A Model Study on a Cobalt Oxide in Selected Electrolytes, Adv. Energy Mater. 11 (2021) 2101737.

https://doi.org/https://doi.org/10.1002/aenm.202101737.

[29] C. Wei, R.R. Rao, J. Peng, B. Huang, I.E.L. Stephens, M. Risch, Z.J. Xu, Y. Shao-Horn, Recommended Practices and Benchmark Activity for Hydrogen and Oxygen Electrocatalysis in Water Splitting and Fuel Cells, Adv. Mater. 31 (2019) 1806296. https://doi.org/10.1002/adma.201806296.

[30] J. Melder, P. Bogdanoff, I. Zaharieva, S. Fiechter, H. Dau, P. Kurz, Water-Oxidation Electrocatalysis by Manganese Oxides: Syntheses, Electrode Preparations, Electrolytes and Two Fundamental Questions, Zeitschrift Fur Phys. Chemie. 234 (2020) 925–978. https://doi.org/10.1515/zpch-2019-1491.

[31] C.C.L. McCrory, S. Jung, J.C. Peters, T.F. Jaramillo, Benchmarking Heterogeneous Electrocatalysts for the Oxygen Evolution Reaction, J. Am. Chem. Soc. 135 (2013) 16977–16987. https://doi.org/10.1021/ja407115p.

[32] Y. Kuang, M.J. Kenney, Y. Meng, W.H. Hung, Y. Liu, J.E. Huang, R. Prasanna, P. Li, Y. Li, L. Wang, M.C. Lin, M.D. McGehee, X. Sun, H. Dai, Solar-driven, highly sustained splitting of seawater into hydrogen and oxygen fuels, Proc. Natl. Acad. Sci. U. S. A. 116 (2019) 6624–6629. https://doi.org/10.1073/PNAS.1900556116.

[33] S. Geiger, O. Kasian, M. Ledendecker, E. Pizzutilo, A.M. Mingers, W.T. Fu, O. Diaz-Morales, Z. Li, T. Oellers, L. Fruchter, A. Ludwig, K.J.J. Mayrhofer, M.T.M. Koper, S. Cherevko, The stability number as a metric for electrocatalyst stability benchmarking, Nat. Catal. 1 (2018) 508–515. https://doi.org/10.1038/s41929-018-0085-6.

[34] Y.T. Kim, P.P. Lopes, S.A. Park, A.Y. Lee, J. Lim, H. Lee, S. Back, Y. Jung, N. Danilovic, V. Stamenkovic, J. Erlebacher, J. Snyder, N.M. Markovic, Balancing activity, stability and conductivity of nanoporous core-shell iridium/iridium oxide oxygen evolution catalysts, Nat. Commun. 8 (2017) 1–8. https://doi.org/10.1038/s41467-017-01734-7.

[35] R. Frydendal, E.A. Paoli, B.P. Knudsen, B. Wickman, P. Malacrida, I.E.L. Stephens, I. Chorkendorff, Benchmarking the Stability of Oxygen Evolution Reaction Catalysts: The Importance of Monitoring Mass Losses, ChemElectroChem. 1 (2014) 2075–2081. https://doi.org/10.1002/celc.201402262.

[36] K. Obata, R. van de Krol, M. Schwarze, R. Schomäcker, F.F. Abdi, In situ observation of pH change during water splitting in neutral pH conditions: impact of natural convection driven by buoyancy effects, Energy Environ. Sci. 13 (2020) 5104–5116. https://doi.org/10.1039/D0EE01760D.

[37] G.H. Simon, C.S. Kley, B. Roldan Cuenya, Potential-Dependent Morphology of Copper Catalysts During CO2 Electroreduction Revealed by In Situ Atomic Force Microscopy, Angew. Chemie Int. Ed. 60 (2021) 2561–2568. https://doi.org/https://doi.org/10.1002/anie.202010449.

[38] M. Teliska, V.S. Murthi, S. Mukerjee, D.E. Ramaker, Site-Specific vs Specific Adsorption of Anions on Pt and Pt-Based Alloys, J. Phys. Chem. C. 111 (2007) 9267–9274. https://doi.org/10.1021/jp071106k.

[39] L.-A. Stern, L. Feng, F. Song, X. Hu, Ni2P as a Janus catalyst for water splitting: the oxygen evolution activity of Ni2P nanoparticles, Energy Environ. Sci. 8 (2015) 2347–2351. https://doi.org/10.1039/C5EE01155H.

[40] F. Claudel, L. Dubau, G. Berthomé, L. Sola-Hernandez, C. Beauger, L. Piccolo, F. Maillard, Degradation Mechanisms of Oxygen Evolution Reaction Electrocatalysts: A Combined Identical-Location Transmission Electron Microscopy and X-ray Photoelectron

Spectroscopy Study, ACS Catal. 9 (2019) 4688–4698. https://doi.org/10.1021/acscatal.9b00280.

[41] A. Grimaud, K.J. May, C.E. Carlton, Y.L. Lee, M. Risch, W.T. Hong, J. Zhou, Y. Shao-Horn, Double perovskites as a family of highly active catalysts for oxygen evolution in alkaline solution, Nat. Commun. 4 (2013) 2439. https://doi.org/10.1038/ncomms3439.

[42] H.-Y. Wang, S.-F. Hung, H.-Y. Chen, T.-S. Chan, H.M. Chen, B. Liu, In Operando Identification of Geometrical-Site-Dependent Water Oxidation Activity of Spinel Co3O4., J. Am. Chem. Soc. 138 (2016) 36–39. https://doi.org/10.1021/jacs.5b10525.

[43] J.T. Mefford, A.R. Akbashev, M. Kang, C.L. Bentley, W.E. Gent, H.D. Deng, D.H. Alsem, Y.-S. Yu, N.J. Salmon, D.A. Shapiro, P.R. Unwin, W.C. Chueh, Correlative operando microscopy of oxygen evolution electrocatalysts, Nature. 593 (2021) 67–73. https://doi.org/10.1038/s41586-021-03454-x.

[44] G. Li, H. Yu, X. Wang, S. Sun, Y. Li, Z. Shao, B. Yi, Highly effective IrxSn1−xO2 electrocatalysts for oxygen evolution reaction in the solid polymer electrolyte water electrolyser, Phys. Chem. Chem. Phys. 15 (2013) 2858–2866. https://doi.org/10.1039/C2CP44496H.

[45] M. Risch, D.M. Morales, J. Villalobos, D. Antipin, What X-Ray Absorption Spectroscopy Can Tell Us About the Active State of Earth-Abundant Electrocatalysts for the Oxygen Evolution Reaction, Angew. Chemie Int. Ed. (2022) doi: 10.1002/anie.202211949. https://doi.org/10.1002/anie.202211949.

[46] D.M. Morales, J. Villalobos, M.A. Kazakova, J. Xiao, M. Risch, Nafion-Induced Reduction of Manganese and its Impact on the Electrocatalytic Properties of a Highly Active MnFeNi Oxide for Bifunctional Oxygen Conversion, ChemElectroChem. 8 (2021) 2979–2983. https://doi.org/10.1002/celc.202100744.

[47] A.J. Villalobos, D.M. Morales, D. Antipin, R. Golnak, J. Xiao, M. Risch, Stabilization of a Mn-Co oxide during oxygen evolution in alkaline media, ChemElectroChem. (2022) e202200482. https://doi.org/10.1002/celc.202200482.

[48] S. Jung, C.C.L. McCrory, I.M. Ferrer, J.C. Peters, T.F. Jaramillo, Benchmarking nanoparticulate metal oxide electrocatalysts for the alkaline water oxidation reaction, J. Mater. Chem. A. 4 (2016) 3068–3076. https://doi.org/10.1039/C5TA07586F.

[49] C.C.L. McCrory, S. Jung, I.M. Ferrer, S.M. Chatman, J.C. Peters, T.F. Jaramillo, Benchmarking Hydrogen Evolving Reaction and Oxygen Evolving Reaction Electrocatalysts for Solar Water Splitting Devices, J. Am. Chem. Soc. 137 (2015) 4347–4357. https://doi.org/10.1021/ja510442p.

[50] W.T. Hong, M. Risch, K.A. Stoerzinger, A. Grimaud, J. Suntivich, Y. Shao-Horn, Toward the rational design of non-precious transition metal oxides for oxygen electrocatalysis, Energy Environ. Sci. 8 (2015) 1404–1427. https://doi.org/10.1039/C4EE03869J.

[51] E. Fabbri, A. Habereder, K. Waltar, R. Kötz, T.J. Schmidt, Developments and perspectives of oxide-based catalysts for the oxygen evolution reaction, Catal. Sci. Technol. 4 (2014) 3800–3821. https://doi.org/10.1039/C4CY00669K.

[52] A.E. Thorarinsdottir, S.S. Veroneau, D.G. Nocera, Self-healing oxygen evolution catalysts, Nat. Commun. 13 (2022) 1243. https://doi.org/10.1038/s41467-022-28723-9.